\documentclass{an}
\usepackage{graphicx}
\usepackage{times}
\Volume{00}              
\Year{0000}              
\Month{00}               
\Pagespan{000}{000}      

\begin{document}
\lhead[\thepage]{Bunker et al.: CIRPASS on Gemini}
\rhead[Astron. Nachr./AN~{\bf XXX} (2003) X]{\thepage}
\headnote{Astron. Nachr./AN {\bf 32X} (2003) X, XXX--XXX}

\title{Extragalactic Integral Field Spectroscopy on the Gemini Telescopes}

\author{Andrew Bunker$^1$, Joanna Smith$^1$, Ian Parry$^1$, Rob
Sharp$^1$, Andrew Dean$^1$,
Gerry Gilmore$^1$
\and Richard Bower$^2$, Mark Swinbank$^2$, Roger Davies$^{2,3}$
\and R.\ Ben Metcalf$^{1,4}$
\and Richard de Grijs$^{1,5}$}
\institute{Institute of Astronomy, Madingley Road, Cambridge CB3\,0HA, UK \ \ ({\tt email:bunker@ast.cam.ac.uk})
\and Department of Physics, University of Durham, South Road, Durham DH1\,3LE, UK
\and Department of Physics, University of Oxford, NAPL Keble Road, Oxford OX1\,3RH, UK
\and Department of Astronomy, University of California, Santa Cruz, CA95064, USA
\and Department of Physics \& Astronomy, University of Sheffield, Hounsfield Road, Sheffield S3\,7RH, UK}

\date{} 

\abstract{
We have been undertaking a
programme on the Gemini 8-m telescopes to demonstrate the power of
integral field spectroscopy, using the optical GMOS spectrograph, and
the new CIRPASS instrument in the near-infrared. Here we present some
preliminary results from 3D spectroscopy of extra-galactic objects, mapping
the emission lines in a 3CR radio galaxy and in a gravitationally lensed
arc, exploring dark matter sub-structure through observations of
an Einstein Cross gravitational lens, and the star formation time-scales
of young massive clusters in the starburst galaxy NGC\,1140.
}

\maketitle

\section{Introduction}
Integral field spectroscopy is a powerful technique with great potential
for furthering the understanding of galaxy evolution.  By simultaneously
producing spectra at each position over a two dimensional region,
Integral Field Units (IFUs) maximise the amount of information
obtained. The sensitivity of spectroscopy to line emission combined with
the areal coverage of IFUs makes them an excellent tool to map the
spatial variation of metallicity, extinction, velocity and star
formation activity in individual galaxies.

\subsection{Our observations}

We have recently carried out the first integral field spectroscopy of
high redshift galaxies with an 8m telescope. These observations were
part of an international demonstration science programme with Gemini
Observatory lead by the Institute of Astronomy, Cambridge, and the
University of Durham. 

In June 2002 we used the optical Gemini Multi-Object Spectrograph
(GMOS) on Gemini-North in IFU mode (Allington-Smith et al.\ 2002).
GMOS is equipped with a fibre-fed IFU with an $\sim 1000$
lenslet array covering an area of $5''\times 7''$ with
$0.2''$-diameter lenslets and with a wavelength range of
$0.4-1.1\,\mu$m. In August 2002 we commissioned the new near-infrared
IFU CIRPASS (the Cambridge IR Panoramic Survey Spectrograph, Parry et
al.\ 2000) on Gemini-South. CIRPASS is a fibre-fed
spectrograph with a 490 lenslet array covering an area of up to
$5''\times 12''$ with the $0.36''$ lenslet scale (a $0.25''$ lenslet
scale is also available). CIRPASS operates in
the $J$- and $H$-band ($1-1.7\,\mu$m). Since then, CIRPASS has been
offered to the Gemini community for a semester, and used for 20 nights
on Gemini-South in 2003.

For both CIRPASS and GMOS we used medium resolution gratings with a
resolving power of $R\sim 3000$ (100 km/s FWHM) enabling us to
work efficiently between the sky lines: the redshifts of our targets
were chosen to have emission lines in ``clean'' regions of the night sky
spectrum.

The demonstration science programme covered a wide range of targets and
science goals, including the nature of high-redshift damped
Lyman-$\alpha$ absorption systems (see Bunker et al.\ 2001) and the star
formation and kinematics of high-redshift galaxies.
In this brief article, we focus on three specific targets.
Other highlights of the demonstration science programme can be
found on \newline {\tt
http://www.ast.cam.ac.uk/$\sim$bunker/CIRPASS}

\section{The radio galaxy 3C324}

\begin{figure}
\resizebox{\hsize}{!}
{\includegraphics{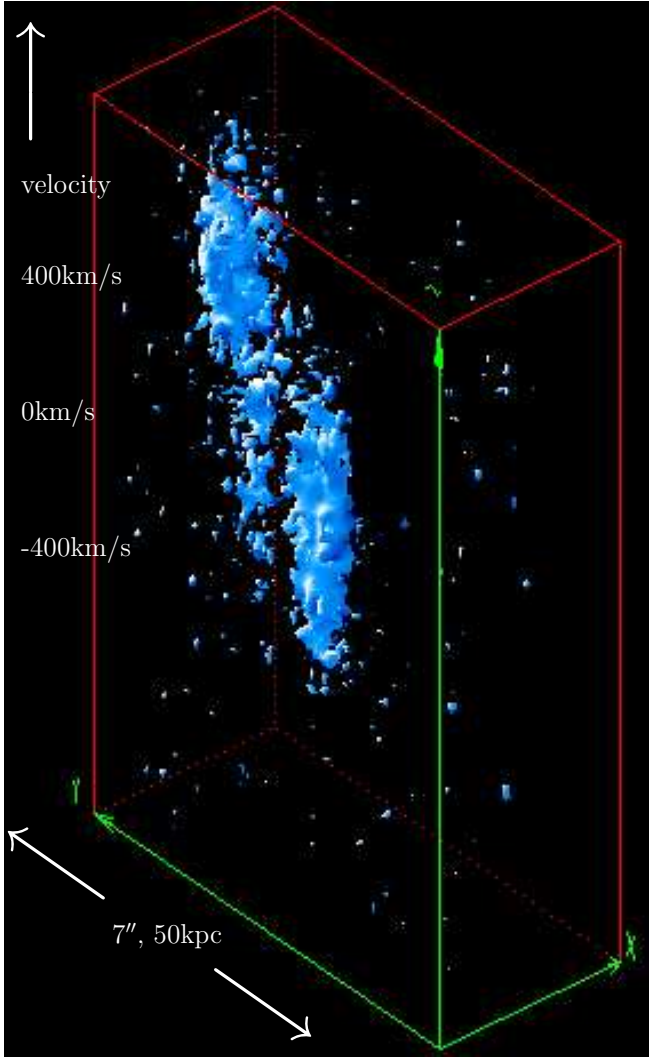}}
\caption{3-D data cube from our GMOS-IFU observations of 3C324 ($z=1.2$),
showing the velocity structure of the [OII]\,3727\,\AA\ line.
The velocity axis is vertical.}
\label{figlabel:3D3C324}
\end{figure}

\begin{figure}
\resizebox{\hsize}{!}
{\includegraphics{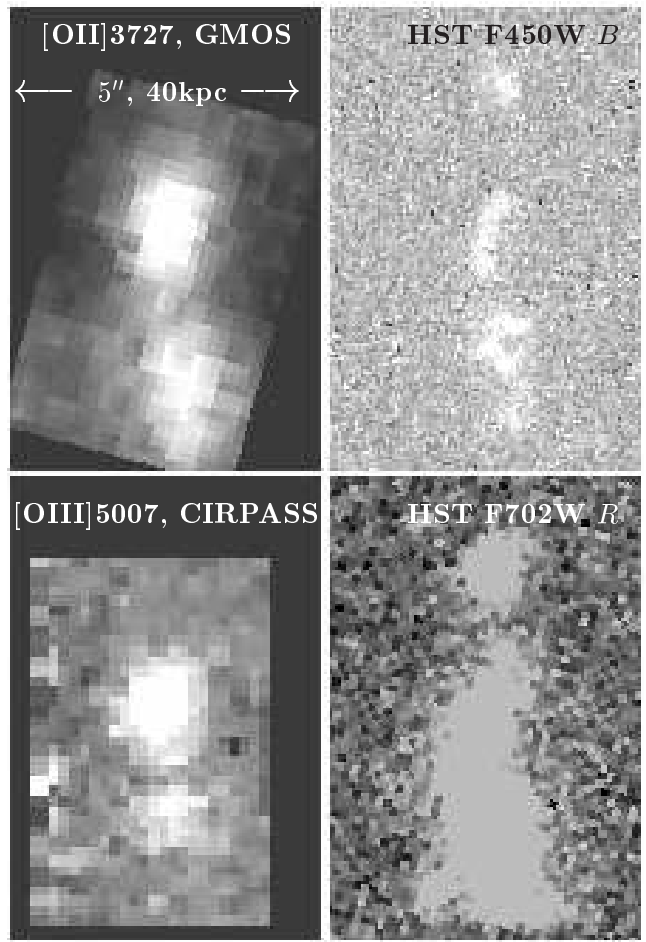}}
\caption{Four images of the $z=1.2$ radio galaxy 3C324, each showing
the same spatial region, 40\,kpc across, with the radio axis
vertical. Clockwise from top left: the [OII]\,3727\,\AA\ line observed
with GMOS-IFU; HST/WFPC2 images in the rest-frame UV image (F450W
$B$-band) and the F702W $R$-band (courtesy of Mark Dickinson); and the
[OIII]\,5007\,\AA\ morphology, obtained through CIRPASS-IFU
spectroscopy in the near-IR. The IFU data has been collapsed in
wavelength to show the spatial distribution of line emission.}
\label{figlabel:3C324}
\end{figure}

We observed the $z=1.206$ radio galaxy 3C324, which has emission line
regions aligned with the radio structure and extending over
$>100$\,kpc (several arcsec, well-matched to the size of the IFUs on
Gemini). Our deep $\sim 2$\,hour spectrum targeted the
[OII]\,3727\,\AA\ emission line using GMOS (Smith et al.\ 2003), and
mapped the spatial variation and velocity structure of the emission
line gas (Fig.~\ref{figlabel:3D3C324}).  The [OII] emission shows two
distinct components of emission line gas with velocities separated by
$\sim$ 800 km/s (as noted by Best et al.\ 2000). The two emission line
components may indicate two separate physical systems undergoing a
merger or radial acceleration by radio jet shocks. We see broad [OII]
emission (FWHM\,$\sim$\,1000\,km/s) indicating large-scale velocity
outflows as well as narrower line emission (FWHM\,$\sim$\,300\,km/s)
in some regions. The velocity spread of the line is sufficient to blur
the [OII] doublet line emission. Using CIRPASS we targeted the
[OIII]\,5007\,\AA\ emission line
(Fig.~\ref{figlabel:3C324}). Comparison with the GMOS observations of
the [OII]\,3727\,\AA\ line yields information on the ionization
mechanisms powering the aligned emission line gas -- the
higher-ionization [OIII] line emission is far more
centrally-concentrated and smaller in spatial extent than the [OII],
perhaps indicating that the central AGN has a greater role in powering
these emission lines than star formation in the individual
high-surface brightness `knots' seen in the rest-UV image (the HST
$B$-band, upper right of Fig.~\ref{figlabel:3C324}).

\section{Spectroscopic Gravitational Lensing and Limits on the Dark
  Matter Substructure in Q2237+0305}

We have used the CIRPASS IFU on Gemini to measure the gravitational
lensing on different size scale of an Einstein cross (the 4--image
quasar Q2237+0305, Huchra et al.\ 1985, Mediavilla et al.\ 1998). In a
project lead by Ben Metcalf, we simultaneously obtain spectroscopy of
the broad line H$\beta$ ($\lambda_{\rm rest}=4861$\,\AA ) and the
forbidden narrow line doublet [OIII]\,4959,5007\,\AA\ from the
$z=1.69$ quasar, lensed by a foreground galaxy at $z=0.03$. These
emission lines arise from physically distinct regions: the narrow line
region around the AGN being much more extended than the broad line
region. Hence, by studying differences in the line ratios in the 4
images, we can assess the significance of sub-structure in the
gravitational lens.

The magnification ratios of the QSO's narrow line region (NLR) and
broad line region (BLR) are found to disagree with each other and with
the published radio and mid-infrared magnification ratios.  The
disagreement between the BLR ratios and the radio/mid-infrared ratios
is interpreted as microlensing by stars in the lens galaxy of the BLR.
This implies that the mid-infrared emission region is larger than the
BLR and the BLR is $< 0.1$\,pc in size.
The disagreement between the
radio/mid-infrared ratios and the NLR ratios is interpreted as a
signature of substructure on a larger scale, possibly the missing
small-scale structure predicted by the standard cold dark matter
(CDM) model.  
%
A substructure mass scale as large as $10^8\,M_{\odot}$ is ruled out
while $10^4\,M_{\odot}$ is too small if the radio and mid-infrared
emission regions have the expected sizes of $\sim 10$\,pc.  The
standard elliptical isothermal lens mass profile is not compatible
with a substructure surface density of $\Sigma_{\rm sub}<
10^8\,M_{\odot}\,{\rm kpc}^{-2}$ at the 95\% confidence level
($\Sigma_{\rm crit}= 10^{10}\,M_{\odot}\,{\rm kpc}^{-2}$ for this
system).  The required substructure surface density at the required
mass scale is high in comparison with the present expectations within
the CDM model.  Lens mass profiles that are flatter than isothermal --
where the surface density in dark matter is higher at the image
positions -- are compatible with smaller quantities of substructure.
The full results are presented in Metcalf, Moustakas, Bunker \& Parry
(2003).

\begin{figure}
\resizebox{\hsize}{!}
{\includegraphics*[80,115][560,335]{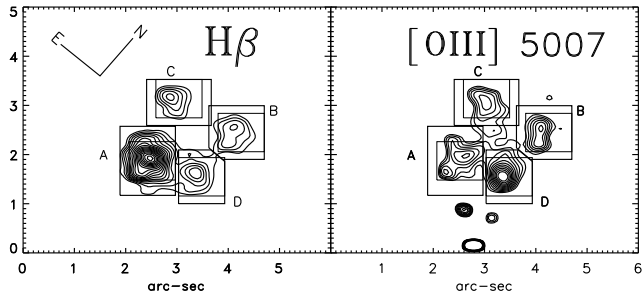}}
\caption{The CIRPASS-IFU spectra of Einstein cross, Q2237+0305, showing the
broad line region in H$\beta$ emission (left) and the narrow line
region in [OIII]\,5007\,\AA\ (right). The different line ratios may
indicate sub-structure in the matter distribution of the foreground
lensing galaxy. The square box apertures are 5\,pixels (0.9\,arcsec)
across, and are used for determining the magnification ratios of the
four components (see Metcalf et al.\ 2003).}
\label{figlabel:EinsteinCross}
\end{figure}

\section{Massive Star Clusters in NGC\,1140}

Moving to lower redshifts, Richard de Grijs has lead a project with
the CIRPASS IFU to study the central star-burst region of NGC\,1140. Our
$1.45-1.67\,\mu$m wavelength coverage includes the bright
[FeII]\,1.64\,$\mu$m emission line, as well as high-order Brackett
hydrogen lines. While strong [FeII] emission, thought to originate in
the thermal shocks associated with supernova remnants, is found
throughout the galaxy, both Br\,12--4 and Br\,14--4 emission are
predominantly associated with the northern starburst region. The
Brackett lines originate from recombination processes occurring on
smaller scales in (young) HII regions. The time-scale associated with
strong [FeII] line emission implies that most of the recent starburst
event in NGC 1140 was induced in the past 35--55\,Myr.  Based on
the spatial distributions of the [FeII] versus Brackett line emission,
we conclude that a galaxy-wide starburst was induced several tens of
Myr ago, with more recent starburst activity concentrated around the
northern starburst region. This work is detailed in de Grijs et al.\
(2003, submitted to MNRAS).
%

\section{Conclusions}

Over the past year we have undertaken an extensive programme to
demonstrate the power of integral field units, using the Gemini
telescopes. This has included the first use of a near-infrared IFU
(CIRPASS) on an 8-m telescope. In this article we have highlighted
observations of three extragalactic sources -a gravitationally lensed
QSO, a radio galaxy, and a low-redshift starburst- to show the variety
of science accessible to IFUs. Other aspects of this programme include
exploring the redshift evolution of of scaling relations such as the
Tully-Fisher law, through GMOS and CIRPASS IFU observations of line
emission from high-redshift disk galaxies (see Smith et al.\ 2003;
Swinbank et al.\ 2003). Elsewhere in these proceedings, Rob Sharp
describes CIRPASS observations of the nearby galaxy M82, and we have
also embarked on an extensive IFU observing campaign to explore the
nature of the galaxies responsible for the damped Lyman-$\alpha$
absorption systems seen in QSO spectra (see Bunker et al.\
2001). Coupling the light grasp of the latest 8-m telescopes with the
new technology of near-infrared IFUs is opening up a hitherto
unexplored parameter space, providing a valuable tool for
studying galaxy evolution.

\begin{figure}
\resizebox{\hsize}{!}
{\includegraphics*[108,143][502,651]{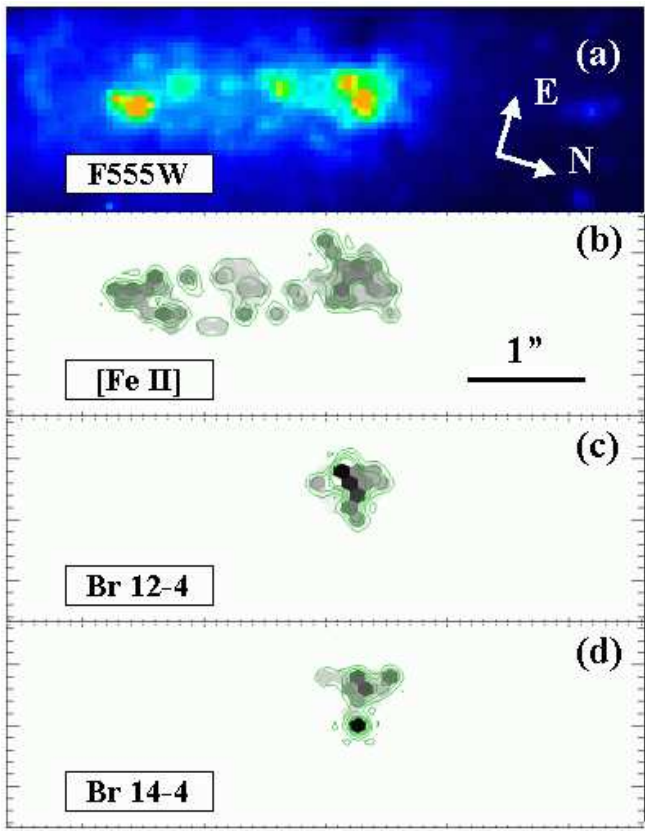}}
\caption{The low-redshift starburst galaxy NGC1140, imaged with
HST/WFPC2 (top panel), and our CIRPASS-IFU spectroscopy at wavelengths
around FeII and hydrogen Brackett lines (lower panels).}
\label{figlabel:ngc1140}
\end{figure}

\acknowledgements This work was done as part of an international
collaboration led by the Institute of Astronomy, Cambridge and the
University of Durham (PIs: Roger Davies, Gerry Gilmore and Andrew
Bunker). We are grateful to the Gemini Science Committee, the GMOS
team and Gemini Observatory for their help and support, in particular
Matt Mountain, Inger J\o rgensen, Jean-Rene Roy, Phil Puxley, Marianne
Takamiya, Bryan Miller and Kathy Roth. We thank
the CIRPASS instrument team, in particular Dave King, Richard McMahon,
Anthony Horton, Steve Medlen and Rachel Johnson. CIRPASS was made
possible by the generous financial support of the Raymond and Beverly
Sackler foundation, and the UK Particle Physics and Astronomy Research
Council. The observations were done as
part of a Director's Discretionary Time programme on Gemini.

\end{document}